\documentclass[twocolumn,showpacs,aps,prl]{revtex4}

\newcommand{\bk}{{\bf k}}

\newcommand{\beq}{\begin{eqnarray}}
\newcommand{\eeq}{\end{eqnarray}}
\newcommand{\beqq}{\begin{eqnarray*}}
\newcommand{\eeqq}{\end{eqnarray*}}

\usepackage{graphicx}
\usepackage{dcolumn}
\usepackage{bm}
\usepackage{color}

\begin{document}

\begin{titlepage}

\title{Electrically Tunable Quantum Spin Hall State in Topological Crystalline Insulator Thin films}

\author{Junwei Liu and Liang Fu}
\address{Department of Physics, Massachusetts Institute of Technology, Cambridge, MA 02139}

\date{\today}

\begin{abstract}

Based on electronic structure calculations and theoretical analysis, we predict the (111) thin films of the SnTe class of three-dimensional (3D) topological crystalline insulators (TCIs) realize the quantum spin Hall phase in a wide range of thickness. The nontrivial topology originates from the inter-surface coupling of the topological surface states of TCIs in the 3D limit.
The inter-surface coupling changes sign and gives rise to topological phase transitions as a function of film thickness. Furthermore, this coupling can be strongly affected by an external electric field, hence the quantum spin Hall phase can be effectively tuned under the experimentally accessible electric field.
\end{abstract}

\pacs{73.21.-b, 73.43.-f, 73.61.-r}

\maketitle

\draft

\vspace{2mm}

\end{titlepage}

Topological crystalline insulators (TCIs) are topological phases of matter protected by  crystal symmetries\cite{fu}.
TCIs have attracted wide attention after the prediction and observation of this phase in IV-VI semiconductors SnTe, Pb$_{1-x}$Sn$_x$Se and Pb$_{1-x}$Sn$_x$Te\cite{hsieh, ando, poland, hasan}. The topological nature of the SnTe class of TCIs arises from  the mirror symmetry of the rocksalt structure, and is manifested in the presence of topological surface states consisting of an even number of Dirac fermions. Recent experiments \cite{vidya,vidya2} have observed that two of the four Dirac fermions on the (001) surface become gapped when the structural distortion takes place and breaks one of the two mirror planes, as predicted by theory\cite{hsieh, serbyn}. This observation establishes the topological protection by crystal symmetry---the defining property of TCIs.

There is currently an intensive experimental investigation of  IV-VI TCIs in the low-dimensional and nanostructure form\cite{taskin,chenh,hguo,Judy1,Judy2,Assaf,Safdar,Sasaki}. Recently we predicted that (001) films of TCIs with an odd number of atomic layers can possess spin-filtered gapless edge states. Interestingly, the edges states can be gapped by a perpendicular electric field that breaks the mirror symmetry. Based on this novel functionality, we proposed a topological transistor device, in which spin and charge transport are coupled and can be simultaneously controlled by the electric field effect\cite{jw_tf}.

Due to the cubic structure in the bulk, electronic structures of IV-VI semiconductor thin films depend crucially on the growth direction\cite{pbte,qian,Ozawa,Ezawa}.
In this work, we study (111) thin films of IV-VI TCIs, which have been grown epitaxially in recent experiments\cite{chenh,taskin}.
We find that (111) thin films can host the quantum spin Hall state\cite{km, bernevig, molenkamp} in a wide range of film thickness with both an even or odd number of layers (despite the different crystal  symmetries in the two cases). The largest nontrivial gap is found to be around 50 meV. Here the quantum spin Hall phase arises from the coherent coupling between the top and bottom topological surface states of TCIs in the 3D limit. This inter-surface coupling mechanism has been previously proposed in model studies of Bi$_2$Se$_3$ thin films \cite{cxliu, Jacob, hlu}, which inspired part of this work. 
Furthermore, we find that the inter-surface coupling in (111) TCI thin films  can be strongly tuned by a moderate external electric field, leading to an electrical control of the topological property.

For convenience,  we will only show below the calculations for SnTe as a representative of TCIs in IV-VI semiconductors. All the qualitative conclusions apply to Pb$_{1-x}$Sn$_x$Te and Pb$_{1-x}$Sn$_x$Se as well. The rocksalt structure of SnTe has two inequivalent (111) surfaces, ending at the plane of Sn or Te atoms respectively.
Both surface states consist of four Dirac cones centered at time-reversal-invariant momenta: one is at $\Gamma$ point and the other three are at $M$ points\cite{jw_kp,jf,hsin,safaei}, as observed in recent angle-resolved photoemission spectroscopy experiments\cite{ando111,polley,chenh}. Different from (001) surface states, these four Dirac cones here are not all equivalent. Specifically, the Dirac point at $\Gamma$ point is well inside the bulk gap while the Dirac points at $M$ points are very close to the valence or conduction band edges. Later, we will show this feature has important consequences for the electronic structures and band topology of (111) thin films.

\begin{figure}[tbp]
\includegraphics[width=7cm]{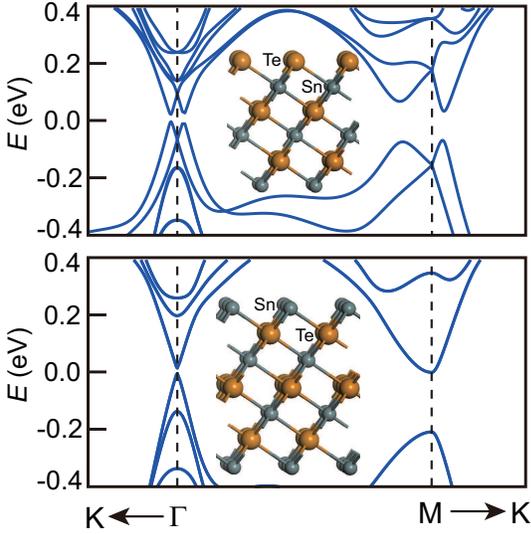}
\caption{The typical band structures of (111) thin films: 24-layer (top) and 25-layer (bottom). The different symmetries of even and odd number of layers thin films (see insets) leads to their qualitatively different band structures.}
\end{figure}

In a thin film, the top and bottom surface states hybridize to open up energy gaps at these four Dirac points. It is important to treat films with an odd and even number of layers separately due to the difference in crystal symmetry: the former has inversion symmetry and the latter doesn't. Our calculations based on the tight-binding model of Lent {\it et al} \cite{lent} show different band structures for the two cases, as shown in Fig.1. Nonetheless, both are semiconductors with a small fundamental gap located at or near the $\Gamma$ point, and a much larger band gap at $M$.
This is because the penetration length of surface states at $M$ is much larger than the one at $\Gamma$, hence the hybridization-induced gap at $M$ is correspondingly larger than the one at $\Gamma$ by orders of magnitude.
As a result of this inequivalence between $\Gamma$ and $M$, (111) films of TCIs in the SnTe class are  narrow-gap semiconductors,
whose properties are controlled by low-energy physics at the $\Gamma$ point.

\begin{figure}[tbp]
\includegraphics[width=7cm]{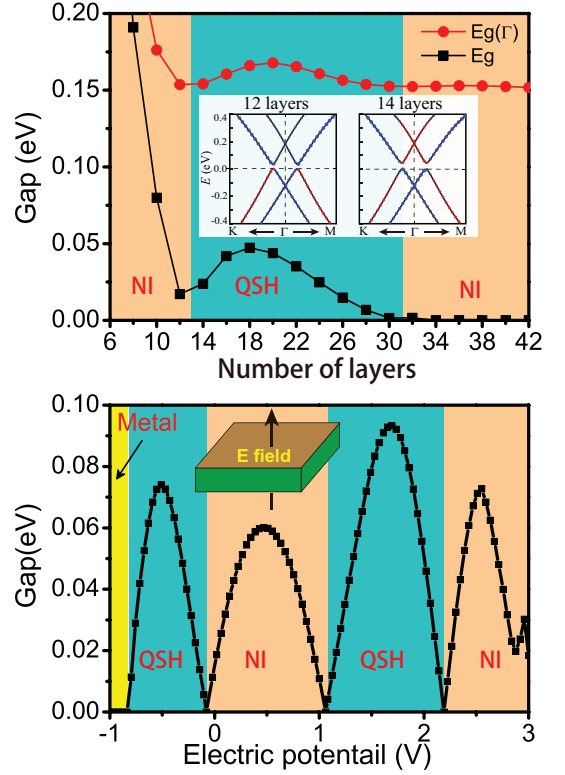}
\caption{The effect of thickness (top) and the effect of external perpendicular electric field on 12-layer thin film (bottom). The quantum spin Hall (QSH) phase can be achieved by changing the thickness or tuning the external electric field. The red/blue dots in the inset of top panel represent that the states mainly compose of Te or Sn atoms and the size means the corresponding weight.}
\end{figure}

To deduce the band  topology, we build an effective $k\cdot p$ Hamiltonian at $\Gamma$. First considering the case of an even number of layers, here the top and bottom surface are different, terminated by Sn-atom and Te-atom respectively, which breaks the inversion symmetry (see insets in Fig.1). Each surface is described by a two-dimensional Dirac fermion: $H_{t, b}= v_{t,b} (k_x s_y - k_y s_x) + E_{t,b} $, where $v_t$ and $v_b$ denote the Dirac velocities of the top and bottom surface respectively; $E_t$ and $E_b$ denote the Dirac point energies of the two surfaces. Denoting the two surfaces by $\tau_z = \pm 1$, the hybridization term to zeroth order in $\bk$ is given by $\tau_x$. This leads to a four-band $k\cdot p$ Hamiltonian for the film:
\beq
H(\bk)=v(k_x s_y-k_y s_x)\tau_z + \delta \tau_z + m \tau_x \label{h}
\eeq
Here for simplicity we have assumed that $v_t = -v_b = v$: including two velocities of different magnitude will not change the conclusion of band topology determined from (\ref{h}).
$\delta=(E_t - E_b)/2$ measures the energy difference of the Dirac points at top and bottom surface, which can be affected by a perpendicular electric field.

By diagonalizing the $H(\bk)$, we obtain four bands with energy-momentum dispersion $E_V(\bk)$, $E_W(\bk)$, $-E_V(\bk)$ and $-E_W(\bk)$, where $E_{V,W}(\bk)$ are given by
\beq
E_{V,W}(\bk)= \sqrt{m^2 + (\sqrt{(k_x^2 + k_y^2) v^2} \pm \delta)^2} \label{disp}
\eeq
The $\delta$-term causes the two degenerate Dirac cones split in energy. For $m=0$, the hole-band of higher Dirac cone and the electron-band of lower Dirac cone will cross each other at $E=0$, forming an circular Fermi surface defined by $(k_x^2 + k_y^2) v^2 = \delta^2$. Now turning on the hybridization term $m \neq 0$ will
open a gap at this circular Fermi surface: $E_g=2|m|$.
In addition, the gap at $\Gamma$ point increases to $E_g(\Gamma)=2\sqrt{m^2+\delta^2}$.
The resulting band structure is  particle-hole-symmetric and has Mexican-hat dispersion, with degenerate band gaps around a circle in $\bk$ space.
This is indeed found in our band structure calculation shown in the top panel of Fig. 1.

Importantly, the band structure (\ref{disp}) shows that the band gap does not close when $\delta$ is tuned to zero, as long as $m$ stays finite.
Therefore, the $Z_2$ topology can be determined by setting $\delta=0$, and is controlled entirely by $m$, the coupling between top and bottom surfaces.
The Hamiltonian for $\delta=0$ takes the form of a two-dimensional massive Dirac Hamiltonian, as in Bernvig-Hughes-Zhang model for HgTe/CdTe quantum well\cite{bernevig}. The sign of $m$ controls the $Z_2$ topology.
Similar to previous proposals for Bi$_2$Se$_3$ thin films\cite{cxliu, Jacob, hlu},  we expect that the inter-surface hybridization $m$, both its magnitude and sign, should depend on the film thickness, thereby leading to gap change and topological phase transition as a function of film thickness.

\begin{figure}[tbp]
\includegraphics[width=8.0cm]{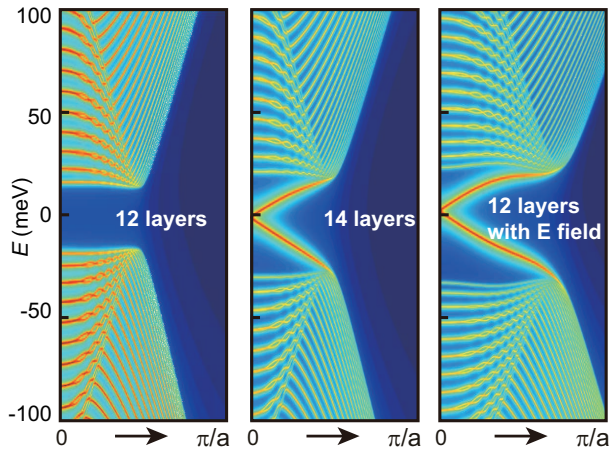}
\caption{The edge states of 12-layers thin film(left panel), 14-layer thin film(middle panel) and 12-layer thin films under electric field (right panel). For 14-layer thin film, there are gapless states with linear dispersion exist in the bulk gap, which confirm that 14-layer thin film is in the quantum spin Hall (QSH) phase. Such kind of edge states are absent for 12-layer thin film, a normal insulator (NI), while we can drive 12-layer thin film into QSH phase by external electric field. (The electric potential between Te-terminated and Sn terminated surfaces is -0.2 V)}
\end{figure}

This is confirmed by our tight-binding calculations. As shown in top panel of Fig. 2, both the fundamental gap $E_g$ and the gap at $\Gamma$ point $E_g(\Gamma)$ change with the film thickness. When the thickness deceases, both gaps exhibit a oscillatory behavior, confirming our expectation from the above $k \cdot p$ model.
The gap oscillation is expected to signal sign change of the Dirac mass $m$, leading to $Z_2$ topological phase transitions between 12-layer and 14-layer thin films. Furthermore, as shown in the inset of top panel of Fig. 2, the conduction bands of 12-layer thin film mainly compose of Sn orbitals, while it's mainly from Te orbitals for 14-layer thin film,  suggesting that 12-layer and 14-layer thin films belong to different topological phases.

To determine which thickness range is topological nontrivial, we directly calculate the edge states in an semi-infinite geometry by using the recursive Green's function method\cite{green}. Without loss of generality, we take armchair-type edges as an example. As shown in Fig. 3, for 14-layer thin film, the linear-dispersion gapless states exist in the bulk gap with a two-fold degeneracy at $\Gamma$ point, named Kramer's degeneracy, forming a one-dimensional Dirac point. In comparison, there are no such kind of helical edge states  in the 12-layer thin film. From such a calculation for other film thickness, we establish unambiguously a topological phase diagram shown in Fig. 2. In particular, all the even number of layers thin films between 14 and 30 layers are quantum spin Hall insulators.

In addition to its sensitivity on the thickness, we find that the inter-surface coupling $m$ can also be strongly affected by a perpendicular electric field\cite{xf}.
Here  the effect of an electric field is modeled by a linearly-increasing electrostatic potential across the film thickness direction, leading to a potential difference $\phi$ between the Te- and Sn-terminated surface. We carefully calculate the evolution of the band structure of 12-layer thin film by varying $\phi$ and find the gap exhibits an very interesting oscillated behavior: it indeed closes at some critical values, implying a topological phase transition induced by the electric field. Specifically, when we apply a negative electric field, the gap will first decrease and closes at $\phi=-0.079$V. As the electric field further increases, the gap reopens and achieve the maximum around 75 meV around $\phi=-0.51$V and then goes down. If the electric field is strong enough, the thin film will become to be a metal. It is worth noting that the critical electric field strength for topological phase transition is only around $4 \times 10^7 \rm{V/m}$, which is an order smaller than the typical breakdown field strength for SnTe related thin films\cite{break}.
To further verify this explicitly, we calculated the edge states for 12-layer thin film under different electric field. As shown in the Fig. 3, without electric field, the edge states are fully gapped for the 12-layer thin film, while when the electric field is stronger than $\phi=-0.079$V, there will be topologically protected edge states in the bulk gap (a typical results with $\phi=-0.2$V shown in the right panel). Similar behavior is seen at positive electric fields, with topological phase transitions at $\phi=1.1$V and 2.2V. The corresponding phase diagram is shown in Fig. 2.

\begin{figure}[tbp]
\includegraphics[width=8.0cm]{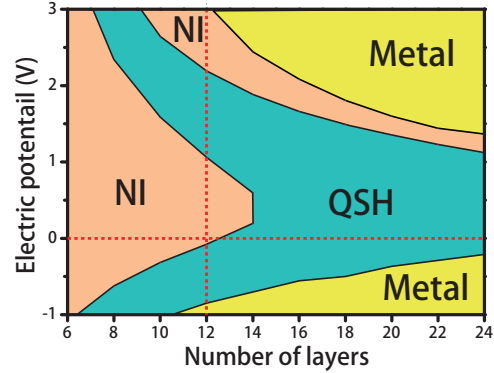}
\caption{The phase diagram for the even number layer of thin films varying with thickness and external electric field.}
\end{figure}

By doing similar calculations for other film thickness, we obtain the complete phase diagram as a function of thickness and electric field, as shown in Fig. 4. From this we find that the quantum spin Hall  phase of thin films thicker than 16 layer is very robust, whereas the thin films between 10-layer and 14-layer are very close to the phase boundary. Topological phase transition can be realized under an experimentally feasible external electric field, which can be set up by using top or bottom gate\cite{xf}.

Now we turn to the case of the odd number of layers thin films, where both surfaces are terminated by the same atoms (Sn or Te atoms) and  inversion symmetry is present (see the insets in Fig. 1). The energy difference term $\delta \tau_z$ in the $k\cdot p$ Hamiltonian (\ref{h}) is no longer allowed, leading to the standard Dirac Hamiltonian $
H=  m \sigma_x + v (k_x s_y - k_y s_x) \sigma_z$.  The presence of inversion symmetry allows us to determine the topological invariant Z2 based on the parity criterion\cite{Z2}.
Clearly, the conduction and valence bands at $\Gamma$ are respectively  the bonding and anti-bonding hybridization of top and bottom surface states, with opposite parities. Tight-binging calculations show that the 2D band structure near $\Gamma$ is described very well by a single-valley Dirac fermion with a small mass, and the dispersion is linear and particle-hole symmetric in a sizable energy range (Fig.1).

\begin{figure}[tbp]
\includegraphics[width=7cm]{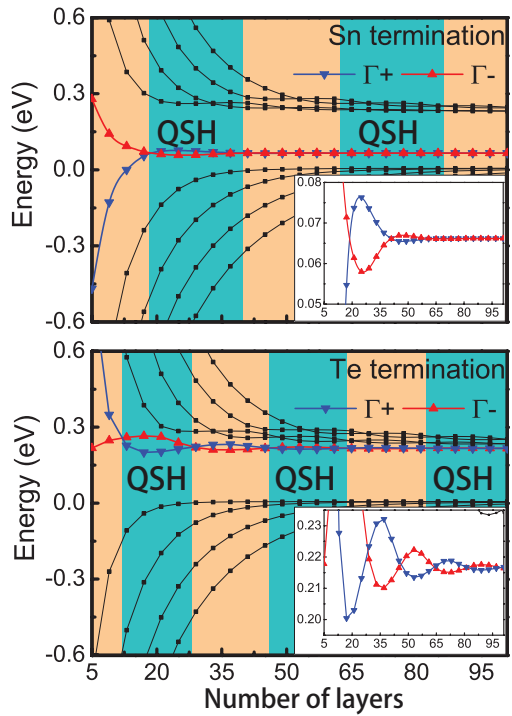}
\caption{The subband energy level diagram at $\Gamma$ point and the topological phase diagram for odd number of layers (111) thin films with Sn-atom (top panel) and Te-atom (bottom panel) terminated. The energy order of $\Gamma_+$ and $\Gamma_-$ states changes in oscillatory behavior with the number of layers, which results in corresponding the topological phase oscillation. The Dirac mass $m$ (gap at $\Gamma$ point) is $m=E(\Gamma_+)-(\Gamma_-)$.}
\end{figure}

Fig.5 shows the energies of subbands at $\Gamma$ for films thinner than 95 layers.
The Dirac mass $m$, given by the energy difference between the bottom conduction and top valence band labelled by $\Gamma_+$ and $\Gamma_-$, depends non-monotonously on the thickness. Importantly, $m$ changes sign at certain critical thickness: 19-, 39-, 63- and 85-layer for Sn termination, and 13-, 27-, 47, 63- and 83-layer for Te termination. The Dirac mass reversal switches the parity of conduction and valence band edges, resulting in topological phase transition between a trivial insulator and a quantum spin Hall insulator. According to our tight-binding calculation, the band gap in the quantum spin Hall phase can reach 18meV at 25 layers for Sn termination, and 66meV at 19 layers for Te termination.
The topological phase transition tuned by film thickness is similar to HgTe/CdTe quantum well\cite{bernevig}. Similar to even number of layers thin films, perpendicular electric potential can also induce topological phase transition for the odd number of layers thin films.

The mechanism for realizing quantum spin Hall phase in TCI thin films via inter-surface coupling is similar to the thin films of topological insulators Bi$_2$Se$_3$ \cite{cxliu, Jacob, hlu}, which partly motivated this work. Due to the much larger penetration length of SnTe surface states, the hybridization gap is much larger than that in Bi$_2$Se$_3$.
Furthermore,  the proposed quantum spin Hall phase in Bi$_2$Se$_3$ was found to depend very sensitively on the film thickness\cite{cxliu, Jacob, hlu}, with $Z_2$ topology changing back and forth with a period of only one or two quintuple layers. In comparison, our results show that  the quantum spin Hall states in TCI films exists in an {\it extended} thickness range and hence are much more robust. Finally, our theoretical analysis of electronic and topological properties of TCI thin films based on inter-surface hybridization may apply to many other TCI materials that are currently being studied\cite{fiete,kindermann,kaisun,dai,A3BO}.

{\it Note added:} After the completion of this work\cite{march}, we learned of an independent work on (111) thin films of TCIs without the external electric field\cite{rb}.

We thank Tim Hsieh for helpful discussions, and R. Buczko for alerting us to his work. This work was supported by the STC Center for Integrated Quantum Materials, NSF Grant No. DMR-1231319.

\bibliographystyle{apsrev}

\begin{thebibliography}{10}


\bibitem{fu}
L. Fu, Phys. Rev. Lett. \textbf{106}, 106802 (2011).

\bibitem{hsieh}
T. H. Hsieh, H. Lin, J. Liu, W. Duan, A. Bansil and L. Fu, Nat. Commun. {\bf 3}, 982 (2012).

\bibitem{ando}
Y. Tanaka, Z. Ren,	T. Sato, K. Nakayama, S. Souma,	T. Takahashi, K. Segawa, Y. Ando, Nat. Phys. {\bf 8}, 800 (2012).

\bibitem{poland}
P. Dziawa, B. J. Kowalski, K. Dybko, R. Buczko	{\it et al.}, Nat. Mater. {\bf 11}, 1023 (2012).

\bibitem{hasan}
S.-Y. Xu, C. Liu, N. Alidoust, M. Neupane, D. Qian  {\it et al.},  Nat. Commun. {\bf 3}, 1192 (2012).


\bibitem{vidya}
Y. Okada, M. Serbyn, H. Lin, D. Walkup {\it et al.}, Science, {\bf 341}, 1496 (2013).


\bibitem{vidya2}
I. Zeljkovic, Y. Okada, M. Serbyn, R. Sankar, D. Walkup, W. Zhou, J. Liu, G. Chang, Y. J. Wang, M. Z. Hasan, F. Chou, H. Lin, A. Bansil, L. Fu, V. Madhavan, arXiv:1403.4906 (2014)

\bibitem{serbyn}
M. Serbyn and L. Fu, Phys. Rev. B {\bf 90}, 035402 (2014).




\bibitem{Safdar}
M. Safdar, Q. Wang , M. Mirza , Z. Wang , K. Xu , and J. He, Nano Lett. {\bf 13}, 5344 (2013)

\bibitem{taskin}
A. A. Taskin, F. Yang, S. Sasaki, K. Segawa, and Y. Ando, Phys. Rev. B {\bf 89}, 121302 (2014)

\bibitem{Judy1}
J. Shen, J. J. Cha, Nanoscale {\bf 6}, 14133 (2014)

\bibitem{Judy2}
J. Shen, Y. Jung, A. S. Disa, F. J. Walker, C. H. Ahn, J. J. Cha, Nano Lett. {\bf 14}, 4183 (2014)

\bibitem{hguo}
H. Guo {\it et al.}, APL Materials {\bf 2}, 056106 (2014)

\bibitem{chenh}
C. Yan {\it et al.}, Phys. Rev. Lett. {\bf 112}, 186801 (2014)

\bibitem{Assaf}
B. A. Assaf, F. Katmis, P. Wei, B. Satpati, Z. Zhang, S. P. Bennett, V. G. Harris, J. S. Moodera and D. Heiman, Appl. Phys. Lett. {\bf 105}, 102108 (2014)

\bibitem{Sasaki}
S. Sasaki, Y. Ando, arXiv:1410.4852 (2014)









\bibitem{jw_tf}
J. Liu, T. H. Hsieh, P. Wei, W. Duan, J. Moodera, L. Fu, Nat. Mat. {\bf 13}, 178 (2014)


\bibitem{pbte}
R. Buczko and L. Cywinski, Phys. Rev. B {\bf 85}, 205319 (2012).

\bibitem{qian}
X. Qian, L. Fu, J. Li, arXiv:1403.3952 (2014)

\bibitem{Ozawa}
H. Ozawa, A. Yamakage, M. Sato, and Y. Tanaka, Phys. Rev. B {\bf 90}, 045309 (2014)

\bibitem{Ezawa}
M. Ezawa, New J. Phys. {\bf 16} 065015 (2014)

\bibitem{km}
C. L. Kane and E. J. Mele, Phys. Rev. Lett. \textbf{95}, 226801 (2005); {\it ibid}, \textbf{95}, 146802 (2005).

\bibitem{bernevig}
B. A. Bernevig, T. L. Hughes and S. C. Zhang, Science \textbf{314},1757 (2006).

\bibitem{molenkamp}
M. K\"onig, S. Wiedmann1, C. Br\"une, A. Roth, H. Buhmann, L. W. Molenkamp, X.-L. Qi, S.-C. Zhang, Science \textbf{318}, 766 (2007).

\bibitem{Jacob}
J.Linder, T. Yokoyama, and A. Sudb$\phi$, Phys. Rev. B {\bf 80}, 205401 (2009).

\bibitem{cxliu}
C. X. Liu, H. J. Zhang, B. Yan, X. L. Qi, T. Frauenheim, X. Dai, Z. Fang, and S. C. Zhang, Phys. Rev. B \textbf{81}, 041307 (2010).

\bibitem{hlu}
H.-Z. Lu, W.-Y. Shan, W. Yao, Q. Niu, and S.-Q. Shen, Phys. Rev. B {\bf 81}, 115407 (2010)


\bibitem{jw_kp}
J. Liu, W. duan and L. Fu, Phys. Rev. B {\bf 88}, 241303(R)(2013)

\bibitem{jf}
J. Wang, J. Liu, Y. Xu, J. Wu, B. L. Gu, W. Duan, Phys. Rev. B {\bf 89}, 125308 (2014)

\bibitem{hsin}
Y. J. Wang, W.-F. Tsai, H.Lin, S.-Y. Xu, M. Neupane, M. Z. Hasan, and A. Bansil, Phys. Rev. B {\bf 87}, 235317 (2013)

\bibitem{safaei}
S. Safaei, P. Kacman, and R. Buczko, Phys. Rev. B {\bf 88}, 045305 (2013)

\bibitem{ando111}
Y. Tanaka, T. Shoman, K. Nakayama, S. Souma, T. Sato, T. Takahashi, M. Novak, K. Segawa, and Y. Ando, Phys. Rev. B {\bf 88}, 235126 (2013)

\bibitem{polley}
C. M. Polley {\it et al.}, Phys. Rev. B {\bf 89}, 075317 (2014)



\bibitem{lent}
C. S. Lent, M. A. Bowen, J. D. Dow, R. S. Allgaier, O. F. Sankey and E. S. Ho, Superlattices Microstruct. {\bf 2}, 491 (1986).

\bibitem{green}
M. P. L. Sancho, J. M. L. Sancho, J. M. L. Sancho and J. Rubio, J. Phys. F: Met. Phys. \textbf{15}, 851 (1985)

\bibitem{xf}
X. Qian, J. Liu, L. Fu and J. Li, Science \textbf{346}, 1344 (2014)

\bibitem{break}
S. Parvanov, V. Vassileva and L. Aljihmania, J. Optoelectron. Adv. Mater. {\bf 7}, 1299 (2005)

\bibitem{Z2}L. Fu and C. L. Kane, Phys. Rev. B \textbf{76}, 045302 (2007).


\bibitem{fiete}
M. Kargarian and G.A. Fiete, Phys. Rev. Lett. {\bf 110}, 156403 (2013).

\bibitem{kindermann}
M. Kindermann, arXiv:1309.1667 (2013).

\bibitem{kaisun}
M. Ye, J. W. Allen, K. Sun,  arXiv:1307.7191 (2013)

\bibitem{dai}
H. Weng, J. Zhao, Z. Wang, Z. Fang, and X. Dai, Phys. Rev. Lett. {\bf 112}, 016403 (2014).

\bibitem{A3BO}
T. H. Hsieh, J. Liu, and L. Fu, Phys. Rev. B {\bf 90}, 081112(2014)






\bibitem{march}
J. Liu, {\it et al.}, http://meetings.aps.org/Meeting/MAR13/
Event/185363 (2013)

\bibitem{rb}
S. Safaei, M. Galicka, P. Kacman and R. Buczko, arXiv:1501.04728 (2015)


\end{thebibliography}

\newpage

\end{document}